\documentclass[prl,superscriptaddress,twocolumn,showpacs,preprintnumbers,%
amsmath,amssymb,floatfix]{revtex4}

\usepackage{graphicx}
\usepackage{dcolumn}
\usepackage{bm}

\begin{document}   
\title{Properties of the phonon-induced pairing interaction
in YBa$_2$Cu$_3$O$_7$ within the local density approximation}

\author{Rolf Heid}
\affiliation{Forschungszentrum Karlsruhe, Institut f\"ur Festk\"orperphysik,
P.O.B. 3640, D-76021 Karlsruhe, Germany}
\author{Roland Zeyher}
\affiliation{Max-Planck-Institut f\"ur Festk\"orperforschung,
             Heisenbergstrasse 1, D-70569 Stuttgart, Germany}
\author{Dirk Manske}
\affiliation{Max-Planck-Institut f\"ur Festk\"orperforschung,
             Heisenbergstrasse 1, D-70569 Stuttgart, Germany}
\author{Klaus-Peter Bohnen}
\affiliation{Forschungszentrum Karlsruhe, Institut f\"ur Festk\"orperphysik,
P.O.B. 3640, D-76021 Karlsruhe, Germany}

\date{\today}

\begin{abstract}
The properties of the phonon-induced interaction between electrons
are studied using the local density approximation (LDA). Restricting
the electron momenta to the Fermi surface we find generally that
this interaction has a pronounced peak for large momentum
transfers and that the interband contributions between bonding and 
antibonding band are of the same magnitude as the intraband ones.
Results are given for various symmetry averages of this interaction
over the Fermi surface. In particular, we find that the dimensionless
coupling constant in the $d$-wave channel $\lambda^d$, relevant for 
superconductivity, is only 0.022, i.e., even about ten times smaller than
the small value of the $s$-wave channel. Similarly, the LDA contribution to the
resistivity is about a factor 10 times smaller than the
observed resistivity suggesting that phonons are not the important
low-energy excitations in high-T$_c$ oxides.   
\end{abstract}


\maketitle

\section{\label{sec:intro} Introduction}
The relevance of phonons for the low-energy properties of high-T$_c$
oxides is presently rather controversially discussed. Some experiments
suggest that the charge carriers near the Fermi surface interact only
weakly with phonons. Examples are   
the magnitude and temperature dependence of the resistivity \cite{Ando} 
and the
rather small effects in the phonon spectrum caused in general
by superconductivity
near or below the transition temperature T$_c$  \cite{Pintschovius}.
For instance, the superconductivity-induced softening and width  
of the zone-center buckling mode in YBa$_2$Cu$_3$O$_7$ corresponds
to a rather small electron-phonon (ep) coupling constant
\cite{Friedl}. The recently measured isotope dependence of angle-resolved 
photoemission spectra (ARPES) \cite{Iwasawa} do not indicate any isotope 
dependence
of the total electronic band width for binding energies down to 200 meV.
This shows the inapplicability of a simple polaron picture, 
caused by a strong ep interaction, to 
cuprates because the associated band narrowing would be sensitive to
different isotopes \cite{Alexandrov}.
 
On the other hand, there exist many observations which have been
taken as evidence for a substantial ep interaction in the cuprates.
The measured isotope effects on the transition temperature
\cite{Franck} and the the superfluid density \cite{Keller},
especially in underdoped samples, are two examples. Another example is
the large width and softening observed in bond-stretching 
phonons in hole-doped cuprates in a very small region in $\bf k$-space
\cite{Reznik}. Similarly, the absence of
a quasi-particle peak in strongly underdoped 
Ca$_{2-x}$Na$_{x}$CuO$_{2}$Cl$_2$ \cite{Shen1} as well as the 
doping dependence of a magnetic transition in the 
frequency-dependent conductivity
were attributed to phonons \cite{Gunnarsson,Mishchenko}.
Finally, electronic self-energy effects in the phonon energy region 
have been observed by ARPES \cite{Lanzara} and by scanning tunneling 
microscopy \cite{Lee} and interpreted in terms of a coupling of 
electrons to a bosonic mode. Whether this mode
is related to phonons is presently unclear but if this is the case
the ep coupling and phonons would certainly be important for the low-energy
physics of cuprates.

Unfortunately, there exists for virtually every interpretation of experiments in
favor or disfavor of phonons alternative explanations.
For instance, the large isotope effects on T$_c$ observed in
the underdoped region do not indicate necessarily a large ep coupling
but may be caused by the pseudogap \cite{Dahm}. Similarly, the 
bosons which interact with electrons near the Fermi surface
in the interpretation of ARPES data may be not phonons but, for instance,
spin fluctuations \cite{spinfluct,Dahm2,Manske}. In view of these uncertainties 
it seems useful to investigate the properties of the
ep coupling independently from any interpretation of experiments, i.e.,
from first-principles using the local density approximation
(LDA). To this end we extend  
recent investigations
on phonon-induced electronic self-energy effects \cite{Heid,Giustino}
and study the momentum, frequency dependence and magnitude of the  
phonon-induced interaction between electrons in detail. 
Though our approach is based on the LDA and deals only with the 
stoichiometric case it takes many features
of these systems such as the complicated phonon spectrum and screening
properties self-consistently and realistically into account. 

\section{\label{sec:int} Computation of the phonon-induced 
interaction}

The retarded, phonon-induced electron-electron interaction, multiplied by
-1 for convenience, is given in momentum space by \cite{Mahan},
\begin{equation}
V({\bf k}\nu,{\bf k}+{\bf q}\mu,\omega)= \sum_j
|g_j({\bf k}\nu,{\bf k}+{\bf q}\mu)|^2\frac{\omega_{{\bf q}j}}
{\omega_{{\bf q}j}^2-(\omega+i\eta)^2}.
\label{V}
\end{equation}
$g_j({\bf k}\nu,{\bf k}+{\bf q}\mu)$ denotes the renormalized amplitude for a
transition from the electronic state with momentum $\bf k$ and band
index $\nu$ to the state with momentum ${\bf k}+{\bf q}$ and band index $\mu$
creating (annihilating) a phonon with branch label $j$ and 
momentum ${\bf q}$ $(-{\bf q})$. $\omega_{{\bf q}j}$ 
denote the phonon frequencies and $\eta$ a positive infinitesimal. 
The amplitudes $g_j$ are obtained within the LDA with the efficient
linear-response technique \cite{Baroni,Heid99}. We used
a 36x36x4-mesh for the electronic momentum $\bf k$ within the Brillouin zone
of YBa$_2$Cu$_3$O$_7$, while phonon frequencies and the 
self-consistent electron-phonon potential were calculated for 
transferred momenta $\bf q$ on a coarser 12x12x4-mesh.
Because even for this reduced mesh an exact calculation was numerically
very demanding, we adopted a two-step procedure. First, phonon related
quantities were calculated exactly by linear response on a 4x4x2-mesh.
Details can be found in Refs.~\onlinecite{Bohnen} and ~\onlinecite{Heid}.
These quantities were then approximated on the 12x12x4-mesh by a
Fourier-interpolation technique.

Under the
usual assumptions it is sufficient to restrict the electronic
momenta ${\bf k}$ and ${\bf k}+{\bf q}$ in $V$ to the Fermi surface.
Our momentum meshes allow to put ${\bf k}$ practically right on the
Fermi surface whereas ${\bf k}+{\bf q}$ was chosen as near as possible
to the Fermi surface. Fermi surface averages were calculated including
also a Gaussian for the one-particle energies
with width $\delta$. The Fermi surface in
YBa$_2$Cu$_3$O$_7$ consists essentially of three bands $\nu,\mu=A,B,C$
where $A$,$B$, and $C$ denote the antibonding, bonding and chain band,
respectively. 

\section{\label{sec:results} Results}
\subsection{Momentum dependence}

\begin{figure}[htpb] 
\includegraphics[angle=0,height=3.1cm,clip]{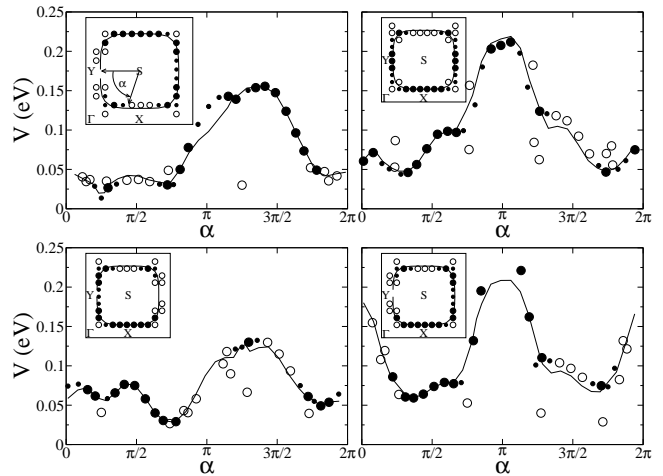}
\hspace{-.2cm}
\includegraphics[angle=0,height=3.03cm,clip]{fig1b.eps}

\includegraphics[angle=0,height=3.1cm,clip]{fig1c.eps}
\hspace{-.2cm}
\includegraphics[angle=0,height=3.03cm,clip]{fig1d.eps}
\caption{\label{fig:1}
Phonon-induced interaction $V({\bf k}B,{\bf k'}B,0)$  (upper part)
and $V({\bf k}A,{\bf k'}B,0)$  (lower part)
for a fixed momentum ${\bf k}$
at the nodal (left diagram) and antinodal (right diagram) Fermi point
as a function of ${\bf k'}$, encircling the Fermi surface around the point
$S=(\pi,\pi)$ in anticlock direction, starting from the antinodal point
(see upper left inset).
Insets show the considered mesh points near the Fermi surface described 
by the solid line.
}
\end{figure}

The upper panels of Fig.~\ref{fig:1} show the intraband contribution 
$\nu=\mu=B$ 
to the static interaction $V({\bf k}\nu,{\bf k'}\mu,0)$. We used 
$k_z=0.125$ and $q_z=0$, i.e., all electronic momenta
are restricted to the plane $k_z=0.125$. The left part in this figure
refers to a fixed momentum $\bf k$ at the nodal, the right part at
the antinodal Fermi point. Symbols indicate electronic states
whose energies differ from the Fermi energy by less than a certain
threshold energy.
The large filled circles refer to 0.1 eV, the small filled
circles to 0.2 eV, the open circles to 0.3 eV.
Lines approximate the Fermi surface and are obtained by a linear
interpolation between neighboring mesh points.
The position in ${\bf k}$-space of the mesh points corresponding to
the different circles is shown in the insets together with the Fermi
line depicted by a solid line. 
$V$ is presented as a function of the
angle $\alpha$ at $S=(\pi,\pi)$ 
between the vectors $({\bf k_x'}-\pi,{\bf k_y'}-\pi)$ and 
$(-\pi,0)$. Varying $\alpha$ between $0$ and $2\pi$ means that ${\bf k'}$,
seen from the point $S$,
moves around the Fermi surface in the anticlockwise sense starting from the
antinodal point $\sim (0,\pi)$. The angle $\alpha$ is illustrated in
the left inset. It is rather straightforward to construct
a continuous curve for $V$ from the discrete points discarding only a very few
points which correspond to momenta ${\bf k'}$ rather far away from the Fermi
surface and thus should be omitted.

\begin{figure} 
\includegraphics[angle=0,height=3.1cm,clip]{fig2a.eps}
\hspace{-.2cm}
\includegraphics[angle=0,height=3.03cm,clip]{fig2b.eps}

\includegraphics[angle=0,height=3.1cm,clip]{fig2c.eps}
\hspace{-.2cm}
\includegraphics[angle=0,height=3.03cm,clip]{fig2d.eps}
\caption{\label{fig:2}
Phonon-induced interaction $V({\bf k}A,{\bf k'}A,0)$ (upper part)
and $V({\bf k}B,{\bf k'}A,0)$ (lower part) 
for a fixed momentum ${\bf k}$
at the nodal (left diagrams) and antinodal (right diagrams) Fermi point
as a function of ${\bf k'}$, encircling the Fermi surface
similar as in Fig.1. Insets show the considered mesh points near the 
Fermi surface described by the solid lines.
}
\end{figure}
\begin{figure}
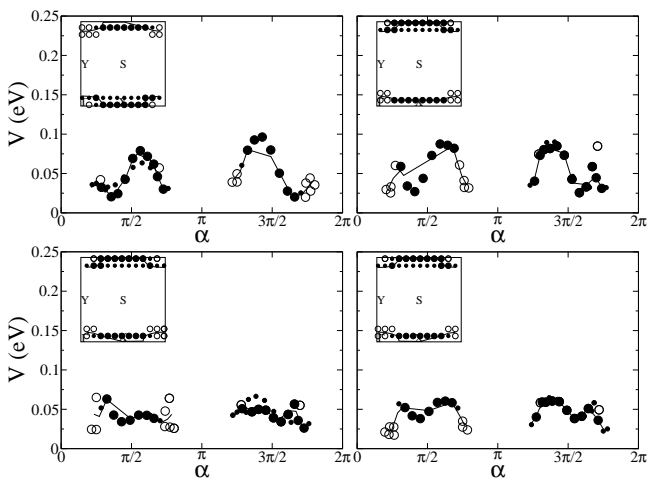
 
\includegraphics[angle=0,height=3.1cm,clip]{fig3a.eps}
\hspace{-.2cm}
\includegraphics[angle=0,height=3.03cm,clip]{fig3b.eps}

\includegraphics[angle=0,height=3.1cm,clip]{fig3c.eps}
\hspace{-.2cm}
\includegraphics[angle=0,height=3.03cm,clip]{fig3d.eps}
\caption{\label{fig:3}
Phonon-induced interaction $V({\bf k}B,{\bf k'}C,0)$ (upper part)
and $V({\bf k}A,{\bf k'}C,0)$ (lower part) 
for a fixed momentum ${\bf k}$
at the nodal (left diagrams) and antinodal (right diagrams) Fermi point
as a function of ${\bf k'}$, encircling the Fermi surface
similar as in Fig.1. Insets show the considered mesh points near the 
Fermi surface described by the solid lines.
}
\end{figure}
The right potential curve in Fig.~\ref{fig:1} should be symmetric with respect
to $\alpha = \pi$ due to the orthorhombic symmetry which is roughly 
fulfilled for our discrete mesh. 
If tetragonal symmetry would apply the left potential curve should be
symmetric with respect to $\alpha = 5\pi/4$ which holds approximately.
Finally, if $V$ depends only on the transferred momentum ${\bf q}$,
the left potential curve, shifted rigidly by
the angle $\alpha = -\pi/4$, would coincide with the right potential
curve. This is qualitatively the case, for instance, the two 
dominating maxima are close to each other after such a shift. 
Quantitatively there are, however,
differences, for instance, the heights of the maxima
differ by about 30\% and the (small) values at zero momentum
transfer by about a factor 3 after the shift. Remarkable is the large 
variation of $V$
by about a factor of 5 between small and large momentum transfers, 
i.e., between
$\alpha =0$ and $\pi$ in the case of the right potential curve. Interesting
for the relation between state and transport relaxation times 
is that $V$ is larger for large than for small  
momentum transfers.  

The lower panels in Fig.~\ref{fig:1} show potential curves for interband
scattering between the antibonding and bonding band. The curves
are rather similar to those of the corresponding upper panels 
in this figure. The only qualitative difference occurs for 
$\alpha \sim 0$ in the right-hand panel where a pronounced
forward scattering peak at the point $Y$ emerges.

The upper and lower parts of Fig.~\ref{fig:2} show potential curves 
in the same way as
in Fig.~\ref{fig:1}, but for the $\nu=\mu=A$ and $\nu=B,\mu=A$ contributions,
respectively. The intraband potential of the antibonding band looks similar
as for the bonding band if $\bf k$ is at the antinodal Fermi point
but very different if $\bf k$ is at the nodal point. In the latter case
it is practically independent of $\alpha$ and rather small. This means
that $V$ depends in this case strongly on ${\bf k}$ and not only on the
transferred momentum ${\bf q}$ which may reflect the strong interaction with   
the chain band. The interband contribution between the bonding and antibonding
band, shown in the lower part of Fig.~\ref{fig:2}, looks qualitatively 
similar as
the interband term between the antibonding and the bonding band of 
Fig.~1 exhibiting well-pronounced 
maxima at large momentum transfers. At small momentum transfers electrons
near the nodal direction interact only weakly whereas those in the antinodal
direction develop a second and rather sharp peak in $V$ at $\alpha=0$ 
which, to a lesser
degree, was also present in the intraband contribution of the antibonding
band. The absolute magnitude of $V$ is in all considered cases similar which
means that interband and intraband terms are comparable in magnitude.

Fig.~\ref{fig:3} shows potential curves for interband scattering
between the band A and B, respectively, and the chain band C. 
Due to the geometry of
the chain band the angle $\alpha$ is restricted to some region
around $\pi/2$ and $3\pi/2$. The potentials peak in general around these 
to values but the absolute values are rather small compared, for instance,
to those in Fig.~\ref{fig:1}.  

\subsection{Frequency-dependence and magnitude of
coupling constants}

\begin{figure}[htpb] 
\includegraphics[angle=0,width=7.2cm,clip]{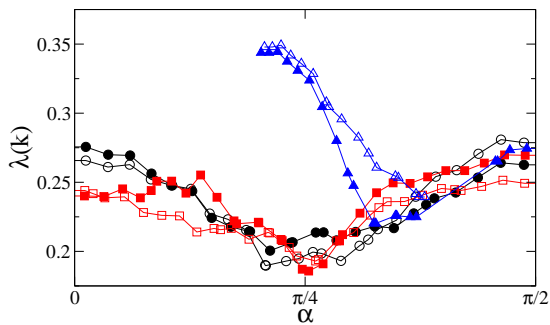}
\caption{\label{fig:4}
(Color online)
Coupling functions $\lambda_A({\bf k})$ (red squares joined by straight lines),
$\lambda_B({\bf k})$ (black circles), and $\lambda_C({\bf k})$ (blue
triangles) for $k_z=0.125$ (filled symbols) and $k_z=0.375$ (empty symbols).
Angle $\alpha$ is the same as in Figs.~1, 2, and 3.
}
\end{figure}

The dimensionless coupling function $\lambda_{\nu}({\bf k})$ \cite{Mahan} can be
written in terms of $V$ as, 
\begin{equation}
\lambda_{\nu}({\bf k}) = 2\sum_{{\bf k'},\mu}V({\bf k}\nu,{\bf k'}\mu,0)
\delta(\epsilon_{{\bf k'}\mu}).
\label{lambda}
\end{equation}
Fig.~\ref{fig:4} shows $\lambda_{\nu}({\bf k})$ for $\nu=A$ (red
squares joined by straight lines), $\nu=B$ (black
circles) and $\nu=C$ (blue triangles) for $\delta = 0.2$ eV as a function
of the angle $\alpha$. 
Using $k_z=0.125$ and $k_z=0.375$ yields two curves denoted  
by filled and empty symbols, respectively, which are close to each other 
illustrating the weak dependence of $\lambda({\bf k})$ on $k_z$.
Both for the bonding and antibonding bands $\lambda({\bf k})$ is
approximately symmetric with respect to $\alpha=\pi/4$ reflecting the
tetragonal symmetry of isolated layers. Moreover, the coupling functions are
smaller along the nodal direction by about 20 to 30 \% compared 
to the antinodal direction which characterizes the anisotropy of 
$\lambda({\bf k})$  
in the layers. The Fermi surface of the chain band starts in our plot
only somewhat below the nodal direction and yields a rather strong and rapidly
varying coupling function $\lambda_C({\bf k})$ reflecting the strong
perturbation of tetragonal symmetry by the chains.

   Dimensionless coupling constants $\lambda_{\nu}^{\alpha}$
with $\alpha=s,s',d,p_x,p_y$ can be defined by
\begin{eqnarray}
\hspace*{-0.5cm} \lambda_{\nu}^\alpha = 
\frac{2}{N_{\nu}^\alpha(0)}\sum_{{\bf k},{\bf k'},\mu} 
V({\bf k}\nu,{\bf k'}\mu,0)\gamma_\alpha({\bf k})\gamma_{\alpha}({\bf k'})
\delta(\epsilon_{{\bf k}\nu})\delta(\epsilon_{{\bf k'}\mu}),
\nonumber \hspace*{-1cm}\\
\label{average}
\end{eqnarray}
with the weight functions $\gamma_s = 1$,
$\gamma_{s'} =  \cos k_x + \cos k_y$, $\gamma_d =
\cos k_x - \cos k_y$, $\gamma_{p_x} = \sin k_x$, and 
$\gamma_{p_y} = \sin k_y$. 
$N_{\nu}^\alpha(0)=\sum_{\bf k} 
\gamma_\alpha^2({\bf k})\delta(\epsilon_{{\bf k}\nu})$ denotes the partial density
of band $\nu$ in the symmetry channel $\alpha$.
The second, third, and fourth
columns in Table I show the calculated values for $\lambda_{\nu}^{\alpha}$
where we approximated the $\delta$-functions in Eq.~(\ref{average})
by Gaussians with width $\delta = 0.2$ eV. The fifth
column contains $\lambda^\alpha$, the sum of the three band
contributions each weighted with the factor $N_{\nu}^\alpha(0)/N^\alpha(0)$ 
where $N^\alpha(0)$ is the total density of electronic states at the 
Fermi surface in the channel $\alpha$. 
These partial density factors enter the total coupling constant
relevant for superconductivity. 

\begin{table}
\caption{Weighted coupling constants.}
\begin{tabular}{c|c|c|c|c}
  & $\lambda_A^\alpha$ & $\lambda_B^\alpha$ & $\lambda_C^\alpha$ & 
$\lambda^\alpha$ \\ \hline 
s & 0.234 &  0.238 & 0.270 & 0.246\\ 
s'& 0.093 & 0.075  & 0.079 & 0.084\\ 
d  & 0.011 & 0.034 & 0.032 & 0.022 \\ 
p$_x$ & -0.020 & -0.042 & -0.022 & -0.028\\ 
p$_y$ & -0.027 & -0.047 & -0.051 & -0.037  
\end{tabular}
\end{table}
  
The numbers in the columns 2-4 of Table I show that the coupling
constants for the three bands are similar in magnitude for each symmetry 
component. The isotropic
$s$-wave component is in all cases substantially larger than those for the 
remaining
``non-trivial'' symmetries. According to Eq.~(\ref{lambda}) 
$\lambda_{\nu}({\bf k})$ and thus also the above coupling constants
contain both intraband ($\mu = \nu$) and interband ($\mu \neq \nu$) 
contributions. Keeping only the
intraband parts diminishes substantially the numbers in the table.
For instance, for the isotropic $s$-wave channel (first line)
the numbers for $A,B,C$ change to 0.090,0.069,0.074,
respectively. This means that interband transitions 
contribute much more to the coupling functions than the 
intraband transitions. Considering $\delta = 0.1$ eV instead of 0.2 eV
does not change much our results: For instance,
the first three numbers in the first line become 0.226,0.229,0.230,
respectively. This suggests that our momentum nets are adequate for
calculating average quantities such as $\lambda_{\nu}^{\alpha}$.

   The most interesting numbers in the table are those in the last
column which determine the phonon contribution to superconductivity.
Isotropic $s$-wave dominates by far but this component 
is ineffective because of the strong Coulomb repulsion. In all other
symmetry channels $\lambda$ is very small. This holds in particular for the
$d$-wave channel where it is only 0.022, i.e., one order of
magnitude smaller than the isotropic $s$-wave value. This value is rather
stable with respect to $\delta$. For instance, it changes 
from 0.022 to 0.027
if $\delta=0.2$ eV is reduced to $\delta =0.1$ eV. Keeping only intraband
contributions one obtains 0.021, i.e., the interband
transitions are negligible in this case.   

\begin{figure}[htpb] 
\includegraphics[angle=0,width=6.2cm,clip]{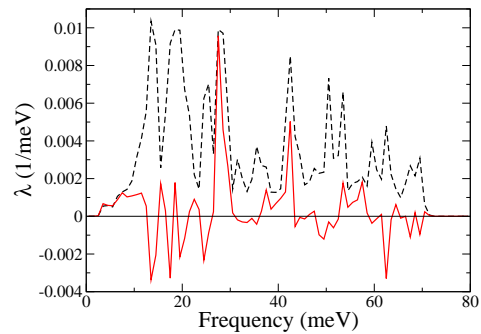}
\caption{\label{fig:5}
(Color online)
Coupling functions $\lambda^s(\omega)$ (black dashed line) and 
$\lambda^d(\omega)$ (red solid line) as a function of frequency using a 
width of 1 meV for the phonon modes. 
}
\end{figure}

Using the Kramers-Kronig transformation for $V$
in Eq.~(\ref{average}) $\lambda_{\nu}^\alpha$ can be written as an integral
from zero to infinity over a frequency-dependent coupling function 
$\lambda_{\nu}^\alpha(\omega)$. The same holds for the total functions 
$\lambda^\alpha$  and $\lambda^\alpha(\omega)$. Fig.~\ref{fig:5} shows
$\lambda^s(\omega)$ (black dashed line) and $\lambda^d(\omega)$ (red
solid line),
using a width of 1 meV for the phonons. 
In the case of $\lambda^s(\omega)$ the phonons give only positive
contributions to $\lambda^s$ which are 
spread out over the whole range of phonon frequencies. The spectrum
is rather peaky because of the usual occurrence of density peaks.
Nevertheless, it is clear that 
it is not possible to attribute the spectrum to a few distinguished 
phonons such as the breathing or buckling modes or bond-stretching
modes. In the case of
$\lambda^d(\omega)$ phonons give both positive and negative contributions
to the spectrum which leads to large cancellations in the integral for 
$\lambda^d$ and thus to a small value for $\lambda^d$. Such cancellations
would be trivial in the case of a momentum-independent $V$ where they would
occur for each phonon frequency separately. In our case $V$ depends
substantially on its two momenta, yet, there are large cancellations,
especially between phonons with different frequencies.
A somewhat curious point is that the high-frequency part of
$\lambda^d(\omega)$ is dominated by two peaks at around 40 and 65 meV
corresponding roughly  to the buckling and breathing phonon frequencies,
respectively. However, these peaks have different signs and thus cancel 
each other to a large extent in $\lambda^d$.
We note that the positive peak in $\lambda^d(\omega)$ at 27 meV is related
to an oxygen buckling-type vibration with odd symmetry with respect to
the CuO$_2$ bilayer. 

\begin{figure}[htpb] 
\includegraphics[angle=0,width=7.0cm,clip]{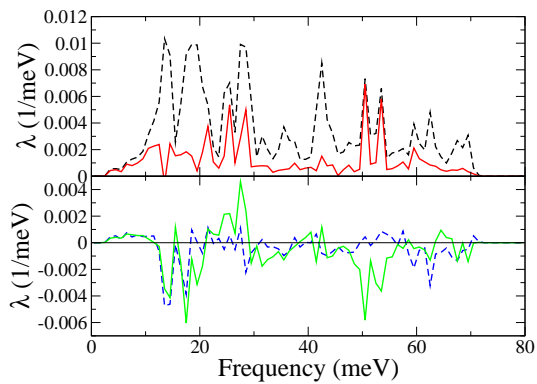}
\caption{\label{fig:6}
(Color online)
Upper panel: Coupling functions $\lambda^s(\omega)$ (black dashed line) and 
$\lambda^{s^\prime}(\omega)$ (red solid line) as a function of frequency using a 
width of 1 meV for the phonon modes. 
Lower panel: Coupling functions $\lambda^{p_x}(\omega)$ (blue dashed line) and 
$\lambda^{p_y}(\omega)$ (green solid line) as a function of frequency using a 
width of 1 meV for the phonon modes. 
}
\end{figure}

In Fig.~\ref{fig:6} we show a comparison between the frequency dependence
of the total coupling functions with $s$ and $s'$ (upper panel)
and with $p_x$ and $p_y$ (lower panel) symmetries. As expected the
$s$-wave coupling function is much larger throughout the whole
frequency interval compared to the other symmetries. Remarkable is
that the couplings with $p_x$ and $p_y$ symmetries are mainly negative
over the whole frequency region and that they differ from each other
reflecting the presence of the chain band and the broken tetragonal
symmetry.   

\subsection{Resisitivity}

\begin{figure}[htpb] 
\includegraphics[angle=0,width=6.2cm,clip]{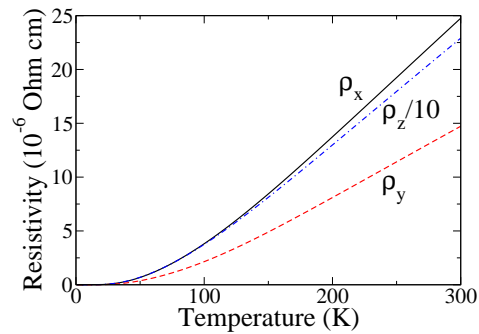}
\caption{\label{fig:7}
(Color online)
Temperature dependence of the resistivity $\rho_i$ along the direction
$i$.
}
\end{figure}

The resistivity in $i$-direction is determined by the transport
coupling constant $\lambda_{i}^{tr}$, given by \cite{Mahan}
\begin{eqnarray}
\lambda_{i}^{tr} = 2 \sum_{{\bf k},{\bf k'},\nu,\mu} 
V({\bf k}\nu,{\bf k'}\mu,0)\delta(\epsilon_{{\bf k}\nu})
\delta(\epsilon_{{\bf k'}\mu}) \cdot \nonumber \\
(v_{i}^2({\bf k}\nu)-v_{i}({\bf k}\nu)v_{i}({\bf k'}\mu))
/\langle v_{i}^2 \rangle,
\label{transport1}
\end{eqnarray}
where $\langle v_{i}^2 \rangle$ denotes the average of 
$v_{i}^2({\bf k}\nu)$ over all pieces of the Fermi surface.
Our calculation yields $\lambda_x^{tr}=0.256$, $\lambda_y^{tr}=0.272$, and
$\lambda_z^{tr}=0.228$. These values are close to $\lambda^s$ illustrating
the fact that the pronounced momentum dependence of V in 
Figs.~\ref{fig:1}-\ref{fig:3} not
necessarily reflects itself in the above coupling constants. Generalizing
$V$ in Eq.~(\ref{transport1}) to a finite  frequency and taking the
imaginary part the right-hand side of this equation is proportional
to the function $\alpha_{tr}^2F(\omega)$ from which the temperature
dependence of the resistivity can be calculated using Eqs.~(4) and (7) 
of Ref. \cite{Allen}. The results are shown in Fig.~\ref{fig:7}.  

With increasing temperature the curves reach between 120-150 K a
quasi-linear regime which approaches in a very smooth way the true linear
region above the highest phonon frequency at around 1000K (not shown
in Fig.~\ref{fig:7}). A typical 
experimental value is $\rho_x(300K) = 290\cdot 10^{-6}$ Ohm cm \cite{Ando}, 
i.e., only
about 10 \% of the experimental scattering is due to phonons. Also the
noticeable deviations from a linear law between T$_c$ = 90K and about 150 K
in the theoretical curves seems to be in conflict with the experiment
for optimally doped samples.

\section{Discussion}

 Our approach treats electronic correlations only in an
approximate way and its applicability to real YBa$_2$Cu$_3$O$_7$ may
be questioned. Unfortunately, the calculation of corrections to our 
results due to
strong electronic correlations is difficult and presently not very
reliable. However, it seems very improbable that our 
main results, namely that $\lambda^d$ and $\lambda^{tr}$ are about 
a factor 50 and 10 too small to account for T$_c$ 
and $\rho$, respectively, will be substantially changed by correlations.
Experimental data on the width of some phonons
are interesting in this respect. The anomalous broadening of the
buckling phonon at the zone center has been quantitatively explained
with LDA results for the ep-interaction, once a large  anharmonic contribution
was subtracted \cite{Friedl}. On the other hand, bond-stretching
phonons at low temperatures show anomalies around the wave vector 
$(0,0.25,0)$ in form
of a sharp and localized  softening and large widths \cite{Pintschovius}
which clearly are beyond the LDA \cite{Heid2}. 
At higher temperatures, where these anomalies are not present, the measured 
widths are still sizable.
Assuming that these widths are entirely due to the ep-interaction
our LDA calculation could only account for about 20 \% of these widths. 
However, part of the widths could also be caused by anharmonicity
like in Ref. \cite{Friedl}. Furthermore, it is not clear how much these
rather localized anomalies in $\bf k$-space contribute to Fermi surface
averaged quantities like $\lambda^{tr}$. 
A large increase of $\lambda^{tr}$ due to correlation-enhanced 
coupling to bond-stretching phonons 
could account for the missing scattering contribution found in the LDA
calculation of $\rho$. On the other hand,
its temperature dependence would be in conflict with
the observed linear temperature between T$_c$ and room temperature 
because the onset of the linear dependence in Fig.~\ref{fig:7} would be 
shifted  to higher
temperatures due to the high frequency of bond-stretching phonons.


\end{document}